\documentclass{article}

\usepackage{oljour}
\usepackage{amssymb}
\usepackage{amsmath}
\usepackage{graphicx}

\usepackage{pdfpages}

\usepackage[colorlinks=true,linkcolor=magenta,citecolor=green,urlcolor=blue]{hyperref}
\makeatletter
\@ifpackageloaded{hyperref}
  {\newcommand{\hrefdoi}[1]{, \href{http://dx.doi.org/#1}{DOI: #1}}}
  {\newcommand{\hrefdoi}[1]{}}
\@ifpackageloaded{hyperref}
  {\newcommand{\hrefarxiv}[2]{, \href{http://arxiv.org/abs/#1}{arXiv:#1 [#2]}}}
  {\newcommand{\hrefarxiv}[2]{}}
\makeatother
\usepackage{url}

\newcommand{\mathid}[1]{{\mathrm{#1}}}
\newcommand{\mathvec}[1]{{\boldsymbol{#1}}}

\begin{document}

\journalname{zp}  % (ZPC)

%\title[de]{}
\title[en]{Ready, set and no action: A static perspective on potential energy surfaces commonly used in gas-surface dynamics}

% Detailed information about all authors of the submission:
% Please use for every author a separate (begin/end) author environment
\begin{author}
  \anumber{1}
  \atitle{MSc}
  \firstname{Vanessa Jane}
  \surname{Bukas}
  \vita{}
  \institute{Technische Universit\"at M\"unchen}
  \street{Lichtenbergstra\ss{}e}
  \number{4}
  \zip{D-85747}
  \town{Garching}
  \country{Germany}
  \tel{+49-89-289-13813}
  \fax{+49-89-289-13622}
  \email{vanessa.bukas@ch.tum.de}
\end{author}

\begin{author}
  \anumber{2}
  \atitle{Dr.}
  \firstname{J\"org}
  \surname{Meyer}
  \vita{}
  \institute{Technische Universit\"at M\"unchen}
  \street{Lichtenbergstra\ss{}e}
  \number{4}
  \zip{D-85747}
  \town{Garching}
  \country{Germany}
  \tel{+49-89-289-13810}
  \fax{+49-89-289-13622}
  \email{joerg.meyer@ch.tum.de}
\end{author}

\begin{author}
  \anumber{3}
  \atitle{Dr.}
  \firstname{Maite}
  \surname{Alducin}
  \vita{}
  \institute{Centro de F{\'i}sica de Materiales CFM, Centro Mixto CSIC-UPV/EHU}
  \street{Paseo Manuel de Lardizabal}
  \number{5}
  \zip{20018}
  \town{San Sebasti{\'a}n}
  \country{Spain}
  \tel{+34-943-01-8418}
  \fax{+34-943-01-5800}
  \email{wapalocm@sq.ehu.es}
\end{author}

\begin{author}
  \anumber{4}
  \atitle{Prof. Dr.}
  \firstname{Karsten}
  \surname{Reuter}
  \vita{}
  \institute{Technische Universit\"at M\"unchen}
  \street{Lichtenbergstra\ss{}e}
  \number{4}
  \zip{D-85747}
  \town{Garching}  
  \country{Germany}
  \tel{+49-89-289-13812}
  \fax{+49-89-289-13622}
  \email{karsten.reuter@ch.tum.de}
\end{author}

\corresponding{joerg.meyer@ch.tum.de}

\abstract{%
In honoring the seminal contribution of Henry Eyring and Michael Polanyi who first introduced the concept of potential energy surfaces (PESs) to describe chemical reactions in gas-phase [Z. Phys. Chem. 12, 279-311, (1931)], this work comes to review and assess state-of-the-art approaches towards first-principle based modeling in the field of gas-surface dynamics. Within the Born-Oppenheimer and frozen surface approximations, the O$_2$-Ag(100) interaction energetics are used as a showcase system to accentuate the complex landscape exhibited by the PESs employed to describe the impingement of diatomics on metal substrates and draw attention to the far-from-trivial task of continuously representing them within all six molecular degrees of freedom. To this end, the same set of \textit{ab initio} reference data obtained within Density Functional Theory (DFT) are continuously represented by two different state-of-the-art high-dimensional approaches, namely the Corrugation-Reducing Procedure and Neural Networks. Exploiting the numerically undemanding nature of the resulting representations, a detailed static evaluation is performed on both PESs based on an extensive global minima search. The latter proved particularly illuminating in revealing representation deficiencies which affect the dynamical picture yet go otherwise unnoticed within the so-called ``divide-and-conquer'' approach.
}
\zusammenfassung{}

% bis zu sechs Schlagwörter (deutsch)
%\schlagwort{}

% Widmung
% dedication
\dedication{}

% Die nachfolgenden Angaben werden in der Regel vom Verlag eingetragen.
% The following information is usually supplied by the publisher.
%======================================================================
\received{}
\accepted{}
\volume{}
\issue{}
\class{}
\Year{}
%======================================================================

\maketitle

\section{Introduction}\label{sec:introduction}

Reflecting the ingenuity of the idea, some 80 years after the seminal contribution of Eyring and Polanyi \cite{Eyring-Polanyi1931} it is difficult to envision chemical reactions without following their footsteps to consider a practical consequence of the Born-Oppenheimer approximation (BOA) \cite{Born1927}: potential energy surfaces (PESs). Knowledge of the potential energy as a function of the atomic positions provides the basis for a most thorough molecular-level understanding of the reaction, in which prominent topological features of the PES like minima and saddle points play a central role. Apart from such static information, a continuous representation of the PES may also be used for explicit dynamical simulations of individual reaction events. If allowed by the numerical efficiency of the latter, appropriate averages over a sufficiently large number of trajectories can routinely be obtained using computational resources commonly available nowadays, thus giving access to the kinetics of the reaction and thereby fulfilling one of the visions of the pioneers Eyring and Polanyi \cite{Eyring-Polanyi1931}.

Although all the more accurate BOA-based numerical solutions of the Schr{\"o}dinger equation have become available during the last eight decades \cite{Nobel1998}, the dimensionality of the problem still remains a major limitation to employing the PES concept in practice. As demonstrated by Eyring and Polanyi \cite{Eyring-Polanyi1931} for simple reactions involving triples and quadruples of atoms in gas phase, a comprehensive understanding of the PES topology can even visually be obtained along two reaction coordinates. For higher dimensions this becomes an increasingly challenging dilemma which research on gas-surface interactions has been battling with for a long time \cite{Kroes2008,book-dgsi-2013}: In contrast to gas-phase reactions, the presence of the solid surface destroys the overall rotational and translational symmetry of the overlaying adsorbate species. Consequently, a description based on intramolecular (in particular) distances alone is no longer possible \cite{Frankcombe2012}, thus inflicting an intrinsically higher dimensionality already for diatomics. In the meantime, the interaction of localized molecular bonds with the extended electronic manifold in particular of catalytically interesting metal surfaces gives rise to complex corrugated PES topologies that cannot readily be described by a reduced number of reaction coordinates. It is therefore not surprising that extensive first-principles based work focusing on prototypical H$_2$ adsorbates has shown that at least all six \emph{molecular} degrees of freedom need to be explicitly considered in order to arrive at a quantitative understanding of central kinetic parameters such as (dissociative) sticking probabilities on rigid surfaces \cite{Gross1997PSS,Gross1998SSR,Kroes2002}.

The exponential growth of computational power has allowed for evaluating thousands of trajectories through molecular dynamics (MD) simulations \cite{McCreery1975,McCreery1977} already some 30 years ago -- when basing the representation of the high-dimensional PES on classical interatomic potentials (CIPs) with ``simple'' analytical forms (like e.g. the Morse potentials used by Eyring and Polanyi \cite{Eyring-Polanyi1931}). Nowadays, kinetic quantities requiring statistical averaging over even orders of magnitude more trajectories can, in fact, readily be obtained for (such) numerically convenient PESs, also e.g. as a function of initial kinetic energy and varying incidence angles \cite{book-dgsi-2013}. It soon became clear, however, that even the semi-empirical CIPs -- with the London-Eyring-Polanyi-Sato (LEPS) potential as the most prominent example -- are severely challenged in capturing the complex bond breaking and making involved in a dissociative adsorption or recombination event \cite{McCreery1977} (although promising steps towards more accurate parametrizations from first principles have recently been reported \cite{book-dgsi-2013,MartinGondre2009,MartinGondre2010}).

On the other hand, \textit{ab initio} electronic structure calculations, even if based on computationally efficient semi-local density-functional theory (DFT), have been far too expensive to allow for an on-the-fly evaluation of the PES. As heralded by first determinations of sticking coefficients through {\em ab initio} molecular dynamics (AIMD) \cite{Gross2007}, this situation might just be changing, at least in cases where less extensive statistical sampling can be accepted. Nevertheless, the huge number of trajectories required to properly determine small reaction probabilities of e.g. $\leq$ $\approx1\%$ \cite{Gross2010} practically excludes direct application of AIMD for many systems for some time still to come. This is particularly true when taking into account nuclear quantum effects, which can be significant for the aforementioned prototypical H$_2$ adsorbates \cite{Gross1998SSR,Kroes1999PSS}.

Given this situation, the state-of-the-art is still defined to its larger extent by a divide-and-conquer type approach. The 6D PES of a diatomic interacting with a frozen surface is first mapped discretely by first-principles calculations and subsequently ``continued in between'' (ensuring differentiability up to at least first order). The actual dynamical simulations are finally conducted on this (numerically undemanding) continuous representation by solving the appropriate classical or quantum mechanical equations of motion. While conceptually appealing, the practical bottleneck of this approach is the difficulty of procuring {\em and} assessing a reliable differentiable interpolation in high (in fact even just six) dimensions that usually are very closely intertwined.

Particularly the latter point shall be the major topic of the present contribution. With a number of high-dimensional interpolation techniques suggested and employed in the context of gas-surface dynamics \cite{Busnengo2000CPL,Busnengo2000JCP,Olsen2002,Crespos2003,Crespos2004,Abufager2007,Gross2006,Lorenz2004,Lorenz2006,Behler2007,Meyer2011,Goikoetxea2012}, we will focus here on the prevalent corrugation-reducing procedure (CRP) \cite{Busnengo2000CPL,Busnengo2000JCP,Olsen2002} and neural networks (NNs) \cite{Witkoskie2005,Behler2011} with proper symmetry adaption \cite{Lorenz2004,Lorenz2006,Behler2007,Goikoetxea2012}. Providing a nice showcase for the PES complexity often met in gas-surface dynamics, we use these two techniques to continuously represent a given DFT data set for O$_2$ at Ag(100) from Alducin {\em et al.} \cite{Alducin2008}. After comparing and highlighting the methodological differences of both approaches in Sect.~\ref{sec:methodology}, we show in Sect.~\ref{sec:results} that the two representations yield notably different reaction probabilities for the oxygen molecule, despite flawlessly passing the typical reliability checks performed to assess the interpolation quality. We therefore perform a thorough investigation into the global energetics of the two representations both in terms of 2D visualizations and of identifying the energetically low lying minima through extensive configurational sampling. This is found a particularly illuminating approach not only in regionally evaluating the accuracy of the PES representations, but also in providing an insightful picture into the intricate PES topologies. For the O$_2$/Ag(100) system it rationalizes the discrepancies found for important dynamical observables within the two interpolation schemes, thereby highlighting the issue of the interpolation reliability and suggesting to generally perform \textit{a priori} static PES examinations within the divide-and-conquer approach.

\section{Methodology}\label{sec:methodology}

\subsection{Coordinate systems}\label{sec:methodology_coords}

\begin{figure}
\includegraphics[width=\colwidth]{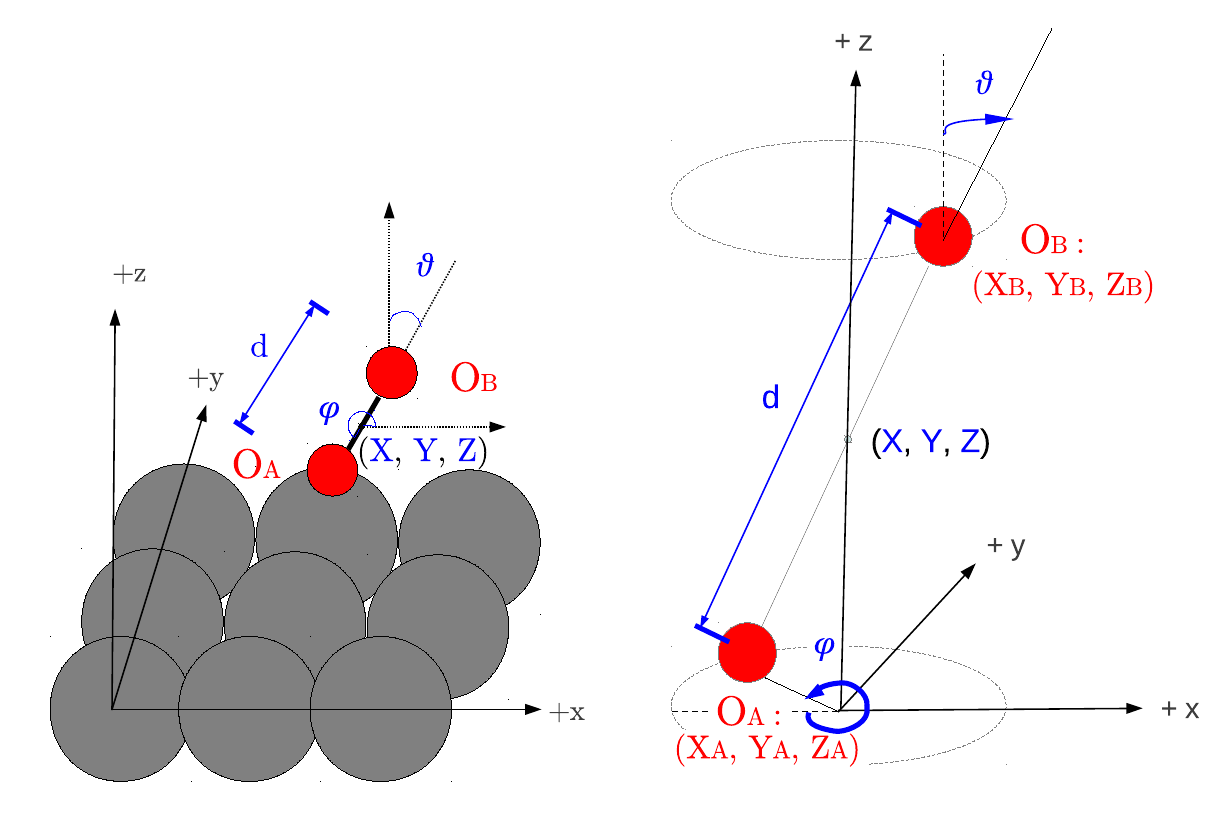}
\caption{Illustration of the 6D coordinate systems used to represent diatomic molecules above a frozen surface. Position and orientation of the molecule can be equivalently described by Cartesian $\mathvec{R}^\mathid{cart} = (X_\mathid{A}, Y_\mathid{A}, Z_\mathid{A}, X_\mathid{B}, Y_\mathid{B}, Z_\mathid{B})$ or spherical $\mathvec{R}^\mathid{sph} = (X, Y, Z, d, \vartheta, \varphi)$ coordinates. The left panel depicts the spherical representation of the adsorbate-surface system with the axes origin located in the surface plane on a top site. The right panel provides a more detailed illustration of the molecular orientation ($\vartheta$, $\varphi$) within the spherical representation and the equivalent Cartesian representation.}
\label{fig:fig1}
\end{figure}

\begin{figure}
\includegraphics[width=\colwidth]{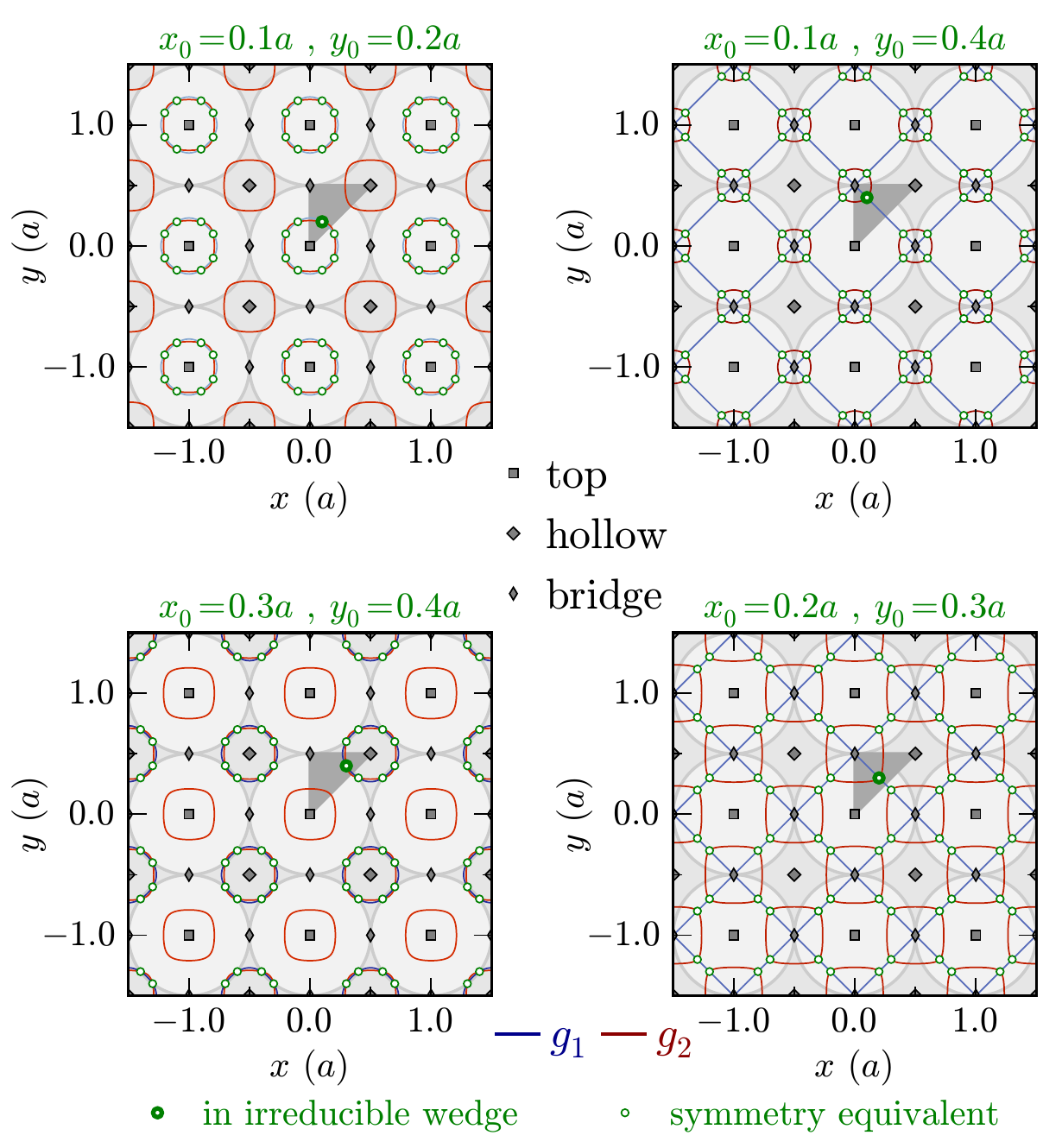}
\caption{Examples for symmetry equivalent lateral coordinates on the Ag(100) surface given in units of the surface lattice constant $a$. For different points $(x_{0},y_{0})$ (thick green circle) in the indicated triangular irreducible wedge (dark gray area), in each panel the equivalents are the intersection points (thin green circles) of contour lines of the functions $g_1$ (blue) and $g_2$ (red) as defined by Eqs.~\ref{eq:g1g2}. Contour values of $g_1$ and $g_2$ are given by $g_1(x_{0},y_{0})$ and $g_2(x_{0},y_{0})$, respectively. In addition to the translation symmetry, note the different local point group symmetry around the high symmetry sites of the surface: top and hollow sites feature four (left panels), whereas the bridge sites feature only two mirror planes perpendicular to the surface (right panels) -- thus resulting in the aforementioned triangular irreducible wedge. Surface atoms are indicated by the large light gray circles, with the one in the center of the plot defining the top site at which the origin is located.}
\label{fig:fig2}
\end{figure}

As indicated in the introduction and depicted in Fig.~\ref{fig:fig1}, the `full-dimensional' representation of the interaction between a diatomic molecule and a rigid substrate surface corresponds to a six-dimensional problem. Straightforwardly, the molecular degrees of freedom can be expressed in terms of the Cartesian coordinates of the two adsorbate constituent atoms A and B
\begin{align}
\label{eq:Rcart}
  \mathvec{R^\mathid{cart}} &
   = ( \underbrace{ X_\mathid{A}, Y_\mathid{A}, Z_\mathid{A} }_{ \mathvec{R_\mathid{A}} }, \, 
         \underbrace{ X_\mathid{B}, Y_\mathid{B}, Z_\mathid{B} }_{ \mathvec{R_\mathid{B}} } ) \quad , \\
\intertext{%
where for all what is to follow, the origin of the coordinate system is located in the top layer of the Ag(100) surface at the position of a silver atom (top site, cf. Figs.~\ref{fig:fig1} and \ref{fig:fig2}). Another representation that more directly conveys equivalent configurations due to surface-induced symmetry as summarized by Fig.~\ref{fig:fig2} is based on spherical coordinates
}
\label{eq:Rsph}
  \mathvec{R}^\mathid{sph} &
    = ( \underbrace{ X, Y, Z}_{ \mathvec{R} } , \, 
          d, \vartheta, \varphi ) \quad ,
\end{align}
with the molecular center of mass $\mathvec{R}$, $d$ the molecule's internuclear distance, and $(\vartheta,\varphi)$ the polar and azimuth angle, respectively, as defined in Fig.~\ref{fig:fig1}. 

While $\mathvec{R}^\mathid{cart}$ can be obtained from $\mathvec{R}^\mathid{sph}$ without any ambiguities 
\begin{equation}
\label{eq:sph2cart}
  \mathvec{R}_{\mathid{A,B}} = \mathvec{R} \, \mp \, \frac{m_{\mathid{B,A}}}{M} \, d \,
    \left( \sin\vartheta \cos\varphi, \sin\vartheta \sin\varphi, \cos\vartheta \right)
    \quad ,
\end{equation}
we emphasize that special care must be taken with properly defining the inverse transformation $\mathvec{R}^\mathid{sph}(\mathvec{R}^\mathid{cart})$:
\begin{subequations}
\label{eq:cart2sph}
\begin{align}
\label{eq:cart2sph_R}
  \mathvec{R} & =  \frac{m_\mathid{A}}{M} \mathvec{R}_\mathid{A} \, + \, \frac{m_\mathid{B}}{M} \mathvec{R}_\mathid{B} \\
\label{eq:cart2sph_d}
  d &= \sqrt{\vphantom{(Z_\mathid{B}-Z_\mathid{A})^2}\smash{
     \underbrace{ {(X_\mathid{B}-X_\mathid{A})}^2 + {(Y_\mathid{B}-Y_\mathid{A})}^2 }_{\substack{\text{$d_{\parallel}^2$}}} +
     \underbrace{ {(Z_\mathid{B}-Z_\mathid{A})}^2 }_{\substack{\text{$d_{\perp}^2$}}} } } \\
\nonumber \\
\label{eq:cart2sph_theta}
  \vartheta &= \arccos{\left(\frac{Z_\mathid{B}-Z_\mathid{A}}{d}\right)} \cdot \frac{180^{\circ}}{\pi} \hspace{33mm} \, \in[0^{\circ},180^{\circ}] \\
\label{eq:cart2sph_phi}
  \varphi &= \mathrm{H}(Y_\mathid{A} - Y_\mathid{B}) \cdot 360^{\circ} \nonumber \\
    &\qquad {} + \mathrm{sign}(Y_\mathid{B}-Y_\mathid{A}) \arccos{\left(\frac{X_\mathid{B}-X_\mathid{A}}{d_{\parallel}}\right)} \cdot \frac{180^{\circ}}{\pi} \quad \in[0^{\circ},360^{\circ}) \quad ,
\end{align}
\end{subequations}
where $m_\mathid{A,B}$ is the mass of atoms A, B and $M=m_\mathid{A}+m_\mathid{B}$ the mass of the entire molecule. The distances $d_{\parallel}$ and $d_{\perp}$ are the components of the internuclear distance $d$ in the $xy$ plane and along the $z$ axis, respectively. Particular care must be taken in defining the validity ranges and reference directions of the angular variables. The convention that has been chosen here (and depicted in Fig.~\ref{fig:fig1}) has the polar angle $\vartheta$ restricted to $[0^{\circ},180^{\circ}]$ as configurations with $\vartheta \in (180^{\circ},360^{\circ})$ may be described by equivalent counterparts in the aforementioned interval by rotating an additional $180^{\circ}$ in $\varphi$. Accordingly, the Heaviside (or step) function $\mathid{H}$ and the sign function are used in Eq.~(\ref{eq:cart2sph_phi}) in order to ensure that configurations in all four quadrants are mapped to the correct value of $\varphi \in [0^{\circ},360^{\circ}]$. Note that permutation symmetry of homo-nuclear diatomics is deliberately not included here, thus distinguishing $\mathvec{R}^\mathid{sph}(\mathvec{R}_\mathid{A},\mathvec{R}_\mathid{B}) \neq \mathvec{R}^\mathid{sph}(\mathvec{R}_\mathid{B},\mathvec{R}_\mathid{A})$. The proper one-to-one mapping established here by Eqs.~(\ref{eq:sph2cart}) and (\ref{eq:cart2sph}) is crucial for the global minima search described in Sect.~\ref{sec:methodology_minima} below, as this involves frequent back-and-forth transformations in order to identify symmetry-equivalent configurations.

In this context and even more so for the neural network interpolation described in Sect.~\ref{sec:methodology_interpolation}, symmetry adapted coordinates defined (for homo-nuclear diatomics) in a similar way to Ref.~\cite{Goikoetxea2012} are of particular importance. These map configurations that are equivalent by the permutation- and surface-induced symmetries (cf. Fig.~\ref{fig:fig2}) to the same $\mathvec{Q}(\mathvec{R}^\mathid{cart}, \mathvec{R}^\mathid{sph}) \in \mathbb{R}^{9}$,
\begin{subequations}
\begin{align}
\label{eq:Q_13}
  Q_{1,3} & \, = \, \frac{1}{2} \sum_{I \in \{A,B\}} \exp( -\tfrac{1}{2} Z_{I} ) \cdot g_{1,2}(X_{i},Y_{i}) \\
\label{eq:Q_24}
  Q_{2,4} & \, = \, \prod_{I \in \{A,B\}} \exp( -\tfrac{1}{2} Z_{I} ) \cdot g_{1,2}(X_{i},Y_{i}) \\
\label{eq:Q_56}
  Q_{5,6} & \, = \, \exp( -\tfrac{1}{2} Z ) \cdot g_{1,2}(X,Y) \\
\label{eq:Q_7}
  Q_7 & \, = \, \exp( -\tfrac{1}{2} Z ) \\
\label{eq:Q_8}
  Q_8 & \, = \, d \\
\label{eq:Q_9}
  Q_9 & \, = \, \left[ \cos(\vartheta) \right]^2
\quad ,
\end{align}
\end{subequations}
where
\begin{subequations}
\label{eq:g1g2}
\begin{align}
\label{eq:g1}
  g_{1}(x,y) & = \frac{1}{4} \left[ \cos \left( \frac{2\pi}{a} x \right) + \cos \left( \frac{2\pi}{a} y \right) \right] + \frac{1}{2} \\
\label{eq:g2}
  g_{2}(x,y) & = \frac{1}{4} \left[ \cos \left( \frac{2\pi}{a} x \right) \cdot \cos \left( \frac{2\pi}{a} y \right) \right] + \frac{1}{2}
\quad,
\end{align}
\end{subequations}
incorporate the periodicity of the square-shaped surface lattice with the lattice constant $a$. As illustrated in Fig.~\ref{fig:fig2}, these have been constructed such that symmetry-equivalent points $(x,y)$ in the surface plane result in the same pair of function values $\left(g_{1}(x,y),\, g_{2}(x,y)\right)$, with a detailed account provided in Ref.~\cite{Meyer2012PhD}.

\subsection{DFT data}\label{sec:methodology_DFT}

Based on the above spherical coordinate system, a discrete set of \textit{ab initio} total-energy data were calculated within DFT by Alducin \textit{et al.} in order to describe the adiabatic interaction of O$_2$ with a (rigid) Ag(100) surface \cite{Alducin2008}. Briefly summarized, the set comprises data for 12 different so-called elbow cuts through the PES, i.e. for molecular configurations of defined lateral position of the O$_2$ center of mass ($X$,$Y$) and defined molecular orientation ($\vartheta$,$\varphi$). Within every elbow, total energies were calculated over a ($Z$,$d$) grid, in which the molecule's distance from the surface $Z$ varies from 0 to 4.5~{\AA} (typically in steps of 0.25~{\AA}) for 10 values of the internuclear distance $d$ (chosen between 0.9 and 2.5~{\AA}). Thus a \textit{regular} $(Z,d)$ grid was constructed containing 2250  energy values of the target PES
\begin{equation}
  V_\mathid{6D}(\mathvec{R}_{i}^\mathid{sph}) = E_{\mathid{O}_2\text{@Ag(100)}}(\mathvec{R}_{i}^\mathid{sph}) - E_\text{Ag(100)} - E_{\mathid{O}_2}
  \quad ,
\end{equation}
and further augmented by values from the molecular O$_2$ binding curve for large distances from the surface $Z \ge 4.5~\mathrm{\AA}$. The \textit{ab initio} total energies of the clean surface ($E_\text{Ag(100)}$), isolated oxygen molecule ($E_{\mathid{O}_2}$) and the interacting system ($E_{\mathid{O}_2\text{@Ag(100)}}$) were obtained within a plane-wave basis set with a cut-off energy of 515 eV and ultrasoft pseudopotentials as implemented in the VASP code \cite{Kresse1993PRB,Kresse1994PRB,Kresse1996PRB,Kresse1996CMS,Vanderbilt1990,Kresse1994JPCM}. The exchange-correlation energy was calculated within the generalized gradient approximation (GGA) due to Perdew and Wang (PW91) \cite{Perdew1992}, and the Ag(100) surface was modeled in a supercell geometry containing a five-layer slab, a $(2 \times 2)$ surface unit-cell, and a vacuum distance of 20.89~{\AA} \cite{Alducin2008}.

\subsection{Continuous PES representations}\label{sec:methodology_interpolation}

Based on the DFT data set described above, two different techniques are employed within the present work to obtain PES representations that can be continuously evaluated along with concomitant forces. The prevalent corrugation-reducing procedure (CRP) \cite{Busnengo2000CPL,Busnengo2000JCP} recognizes that a large part of the difficulties behind the interpolation of a six-dimensional function $V_\mathid{6D}$ describing the PES of a diatomic at a solid surface arises from the large energy variations connected e.g. with the strongly repulsive regions at short molecule-surface distance. Much of this corrugation, however, is equally encountered when the constituent isolated atoms approach the surface. The central idea of CRP lies therefore in decomposing the 6D molecular PES $V_\mathid{6D}$ into a superposition of the 3D atom-surface interactions $\mathcal{R}_\mathid{A,B}$ for both atoms (at their coordinates in the diatomic) and a smoother function $\mathcal{I}^\mathid{CRP}$,
\begin{equation}
\label{eq:CRP}
  V_\mathid{6D}^\mathid{CRP}(\mathvec{R}^\mathid{sph},\mathvec{R}^\mathid{cart}) =
    \mathcal{I}^\mathid{CRP}(\mathvec{R}^\mathid{sph}) + \mathcal{R}_\mathid{A}(\mathvec{R}_\mathid{A}) + \mathcal{R}_\mathid{B}(\mathvec{R}_\mathid{B})
  \quad .
\end{equation}
$\mathcal{I}^\mathid{CRP}$ becomes then a more convenient target for the interpolation. In practice, this obviously requires additional \textit{ab initio} data for the construction of continuous representations of $\mathcal{R}_\mathid{A,B}$, which simplifies to a single function in the present case of a homonuclear diatomic. However, the lower dimensionality of $\mathcal{R}_\mathid{A,B}$ makes this a much simpler task -- including a proper incorporation of symmetry \cite{Busnengo2000JCP}. In this like in many preceding works, $\mathcal{I}^\mathid{CRP}$ is then obtained for an arbitrary configuration $\mathvec{R}^\mathid{sph}$ within the range covered by the discrete set $\{ \mathvec{R}_{i}^\mathid{sph} \}$ for which DFT energies $\{ V_\mathid{6D}( \mathvec{R}_{i}^\mathid{sph} ) \}$ are available. This is typically done by decoupling the six-dimensional problem into four at most two-dimensional independent interpolation steps, which can be schematically summarized as follows:
\begin{enumerate}
\item\begin{math}
  \mathcal{I}_{0}^\mathid{CRP}(\{ \mathvec{R}_{i}^\mathid{sph} \}) 
    \; \xrightarrow{(Z_i,d_i):\,\text{splines}} \; 
  \mathcal{I}_{1}^\mathid{CRP}(Z,d \,;\, \{ X_i,Y_i,\vartheta_i,\varphi_i \})
\end{math}
\item\begin{math}
  \mathcal{I}_{1}^\mathid{CRP}(Z,d \,;\, \{ X_i,Y_i,\vartheta_i,\varphi_i \}) 
    \; \xrightarrow{\varphi_i:\,\text{Fourier}} \; 
  \mathcal{I}_{2}^\mathid{CRP}(Z,d,\varphi \,;\, \{ X_i,Y_i,\vartheta_i \})
\end{math}
\item\begin{math}
  \mathcal{I}_{2}^\mathid{CRP}(Z,d,\varphi \,;\, \{ X_i,Y_i,\vartheta_i \})
    \; \xrightarrow{(X_i,Y_i):\,\text{Fourier}} \; 
  \mathcal{I}_{3}^\mathid{CRP}(X,Y,Z,d,\varphi \,;\, \{ \vartheta_i \})
\end{math}
\item\begin{math}
  \mathcal{I}_{3}^\mathid{CRP}(X,Y,Z,d,\varphi \,;\, \{ \vartheta_i \})
    \; \xrightarrow{\vartheta_i:\,\text{splines}} \; 
  \mathcal{I}_{4}^\mathid{CRP}(X,Y,Z,d,\vartheta,\varphi)
\end{math}\\[1ex]
with $ \mathcal{I}_{4}^\mathid{CRP}(X,Y,Z,d,\vartheta,\varphi) \equiv \mathcal{I}^\mathid{CRP}( \mathvec{R}^\mathid{sph} )$.
\end{enumerate}
The interpolation functions employed in steps two to four are chosen according to generally expected energy variation in these degrees of freedom, and are carefully adapted to its symmetry \cite{Busnengo2000CPL,Busnengo2000JCP}.

Neural networks (NNs) as the second prevalent interpolation approach are a highly flexible, non-linear model originally inspired by neuroscience \cite{Behler2011}. Even with utmost mathematical rigor they are shown to be capable of approximating any PES to -- at least in principle -- arbitrary accuracy \cite{Cybenko1989,Hornik1989}. This motivates to target $V_\mathid{6D}$ directly and equally in its full six dimensionality, allowing to capture intertwinement of degrees of freedom which are treated independently in the prevalent interpolation strategy hitherto employed in the CRP context described above. Further unlike the latter, the input data are not required to fall on regular grids, thus enabling the addition of individual points depending on the desired accuracy in certain PES ``regions'' \cite{Behler2007}. Extrapolation capabilities of NNs beyond the coordinate ranges of the input data can also supersede those of the above described individual interpolations due to the inherent limitations of the underlying fixed analytic forms employed in CRP \cite{Behler2011}. However, this inherent flexibility comes as a mixed blessing: Correct symmetry properties can only be guaranteed by presenting symmetry adapted coordinates to a NN \cite{Behler2007,Goikoetxea2012,Meyer2012PhD}, like those introduced in Sect.~\ref{sec:methodology_coords}. The NN thus acts as function $\mathcal{F}^\mathid{NN}(\mathvec{Q})$ for the continuous PES representation according to
\begin{equation}
\label{eq:NN}
  V_\mathid{6D}^\mathid{NN}(\mathvec{R}^\mathid{cart},\mathvec{R}^\mathid{sph}) = \mathcal{F}^\mathid{NN}(\mathvec{Q}(\mathvec{R}^\mathid{cart},\mathvec{R}^\mathid{sph}))
  \quad .
\end{equation}
In practice and as further detailed in preceding work \cite{Lorenz2006}, various multilayer feed-forward NNs of different topologies are fitted (aka ``\emph{trained}'' in NN lingo) to the DFT input data, after some (166 in the present case) randomly selected entries have been moved from the training into a so-called test set. During the training iterations (aka ``\emph{epochs}''), the latter set is employed to monitor the extrapolation abilities. Using the adaptive extended Kalman filter (EKF) algorithm including the modifications described in \cite{Lorenz2006} for training, the sensitivity of the NN is further focused on the energetically presumably most relevant parts of the PES by assigning training weights to the input data according to
\begin{align}
\label{eq:omegaQ}
  \omega_{\mathvec{Q}_i}(V_\mathid{6D}(\mathvec{R}_i^\mathid{sph})) & =  \alpha \, \exp(-\beta \, V_\mathid{6D}(\mathvec{R}_i^\mathid{sph}))
  \quad .
\end{align}
The parameters $\alpha$ and $\beta$ are chosen in the present work so that Eq.~(\ref{eq:omegaQ}) maps the energy interval of the DFT data $[ -0.25 \mathrm{eV} ; +2.5 \mathrm{eV}]$ to 
\begin{equation*}
  \left[ \frac{1}{\omega_{\mathvec{Q}_i}(+2.5 \mathrm{eV})} ; \frac{1}{\omega_{\mathvec{Q}_i}(-0.25 \mathrm{eV})} \right] = \left[ \frac{1}{500} ; 1\right]. 
\end{equation*}
Out of over 70 different attempts we obtained the best fit presented in Sect.~\ref{sec:results_rmse} below for a $\{9 - 25 - 25 - 1 \, ttl\}$ NN topology (see e.g. Refs.~\cite{Lorenz2006,Behler2007} for details on this notation) and the EKF parameters $\lambda(0)=0.9610$ and $\lambda_0=0.99670$.

\subsection{Dynamical simulations}\label{sec:methodology_dynamics}

The dynamics on both the CRP and NN PESs are analyzed with quasi-classical trajectory calculations that include the initial zero point energy of the O$_2$ molecule. A classical microcanonical distribution of the internuclear distance $d$ and its conjugate momentum $p_d$ is used for the molecule in the quasi-classical equivalent of its ground rovibrational state, which was found equal to 97.5 meV \cite{Alducin2008}. All results are derived from the evaluation of 10,000 trajectories, starting at a distance of $Z = 9\,\mathrm{\AA}$ from the surface, where the PES value for O$_2$ at its equilibrium bond length $d_\mathid{eq} = 1.24\,\mathrm{\AA}$ is zero in both PES representations. 

Along the trajectory calculations, the following reaction events are distinguished:
\begin{enumerate}
\item dissociation, when the molecule's internuclear distance reaches the value $d = 2.4\,\mathrm{\AA}$ ($\approx 2 d_\mathid{eq}$) with a positive radial velocity; 
\item reflection, when the molecular center reaches the initial starting distance of $9\,\mathrm{\AA}$ above the surface and with a positive $Z$-velocity; and 
\item molecular trapping, when the molecule is neither dissociated or reflected after 15 ps. 
\end{enumerate}
This classification is the basis to arrive at the central kinetic parameters analyzed in this work, namely the dissociative sticking coefficient and molecular trapping probability, defined as averages over the obtained trajectory data, i.e. the fraction of correspondingly classified trajectories over the total simulated trajectories for the given initial kinetic energy.

\subsection{Global minima search}\label{sec:methodology_minima}

The low computational cost associated with the evaluation of energies and forces on the continuous PES representations allows for extensive configurational sampling in the search for energetically low lying minima. We thus employ a rather brute-force scheme that involves a large number of individual independent geometry optimizations starting from different initial configurations. In these initial configurations the O$_2$ molecule has random lateral positions within the irreducible wedge of the fcc(100) surface unit-cell (cf. Fig.~\ref{fig:fig2}), random distances from the surface and bond lengths between $1.0-2.5\,\mathrm{\AA}$ and any angular orientation. Each configuration is subject to local geometry optimization within the \textit{Atomic Simulation Environment} (ASE) \cite{Bahn2002} until residual forces on the O atoms fall below $0.001\,\mathrm{eV}/\mathrm{\AA}$. The thus optimized geometries are identified as on-surface adsorption states, if the O$_2$ distance from the surface is positive ($Z>$0) and its bond length falls within the range $[1\,\mathrm{\AA}, 2.5\,\mathrm{\AA}]$. Despite the use of symmetry-reduced initial configurations, minimization paths can easily lead out of the irreducible wedge and thus yield potentially symmetry-equivalent minima. The symmetry-adapted coordinates described in Sect.~\ref{sec:methodology_coords} were correspondingly used to conveniently compare to results from previous iterations and finally establish the database of unique adsorption states discussed below. The rate with which new such states were added to the set was continuously monitored during the sampling run, and used as a criterion to stop the search. After 10,000 geometry optimizations on each PES representation, this criterion suggested that all inequivalent low-energy minima had been identified.

\section{Results \& Discussion}	\label{sec:results}

\subsection{Interpolation errors vs. reaction probabilities}	\label{sec:results_rmse}

\begin{figure}
\includegraphics[width=\colwidth]{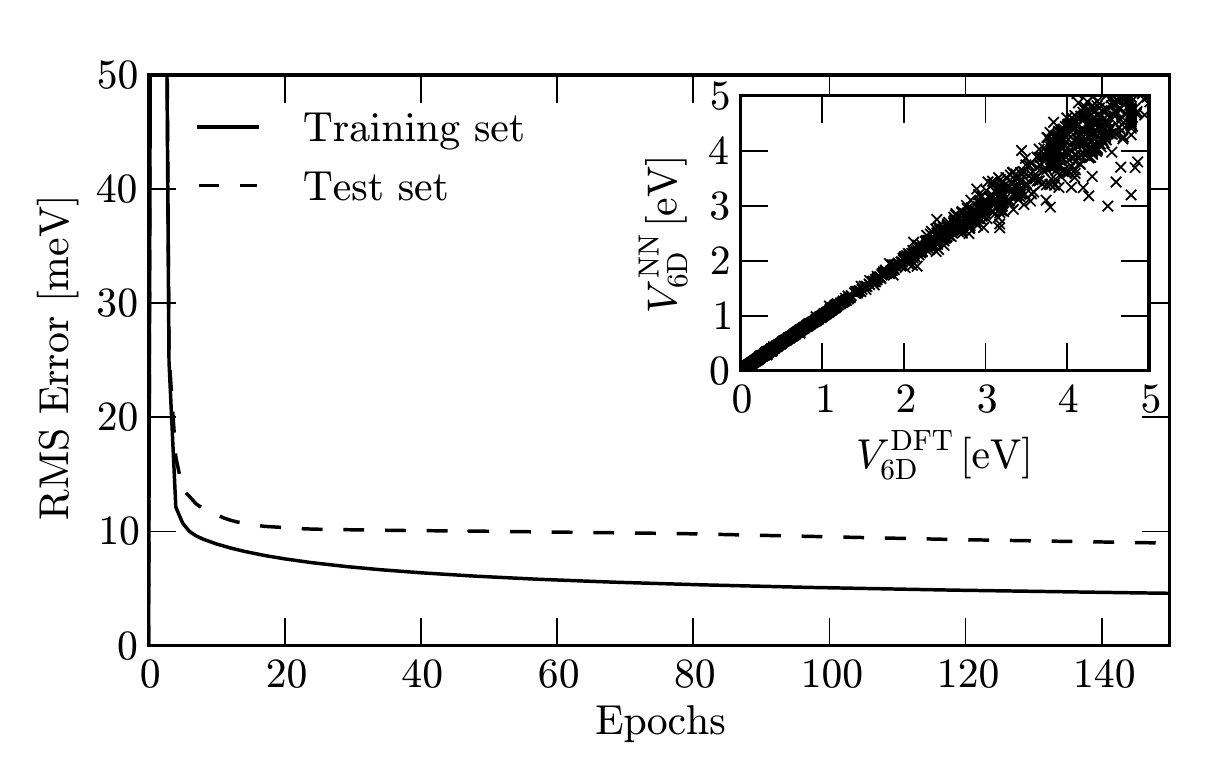}
\caption{Evolution of the training and test root mean square error (RMSE) as a function of the number of \emph{epochs} performed during the Neural Network training. The inset shows the resulting Neural Network energies with respect to the corresponding DFT input data.}
\label{fig:fig3}
\end{figure}

The reliability assessment of a continuous PES representation within the divide-and-conquer approach is hitherto commonly restricted to an evaluation of the interpolation or fitting quality. Within the CRP strategy, the accuracy is correspondingly checked by comparing interpolated values with calculated DFT data not included in the interpolation procedure. In the earlier work of Alducin \textit{et al.} \cite{Alducin2008} such checks indicated errors to fall below 100 meV for a number of different molecular configurations at distances $Z \geq 1.5\,\mathrm{\AA}$. Within the NN approach this kind of performance check is already included in the fitting procedure itself by monitoring the evolution of the root mean square error (RMSE) with each epoch, cf. Sect.~\ref{sec:methodology_interpolation}. For the best fit achieved in the present work the latter is depicted in Fig. \ref{fig:fig3}, arriving at train and test set RMSEs of 4.59 meV and 8.99 meV, respectively, when terminating the NN training after 150 epochs. The inset of Fig. \ref{fig:fig3} shows the resulting deviation of the fitted energies to the reference DFT values, which -- as expected by the choice of the assigned weights -- increases in the more repulsive (and allegedly dynamically less relevant) regions of the PES. 

\begin{figure}
\includegraphics[width=\colwidth]{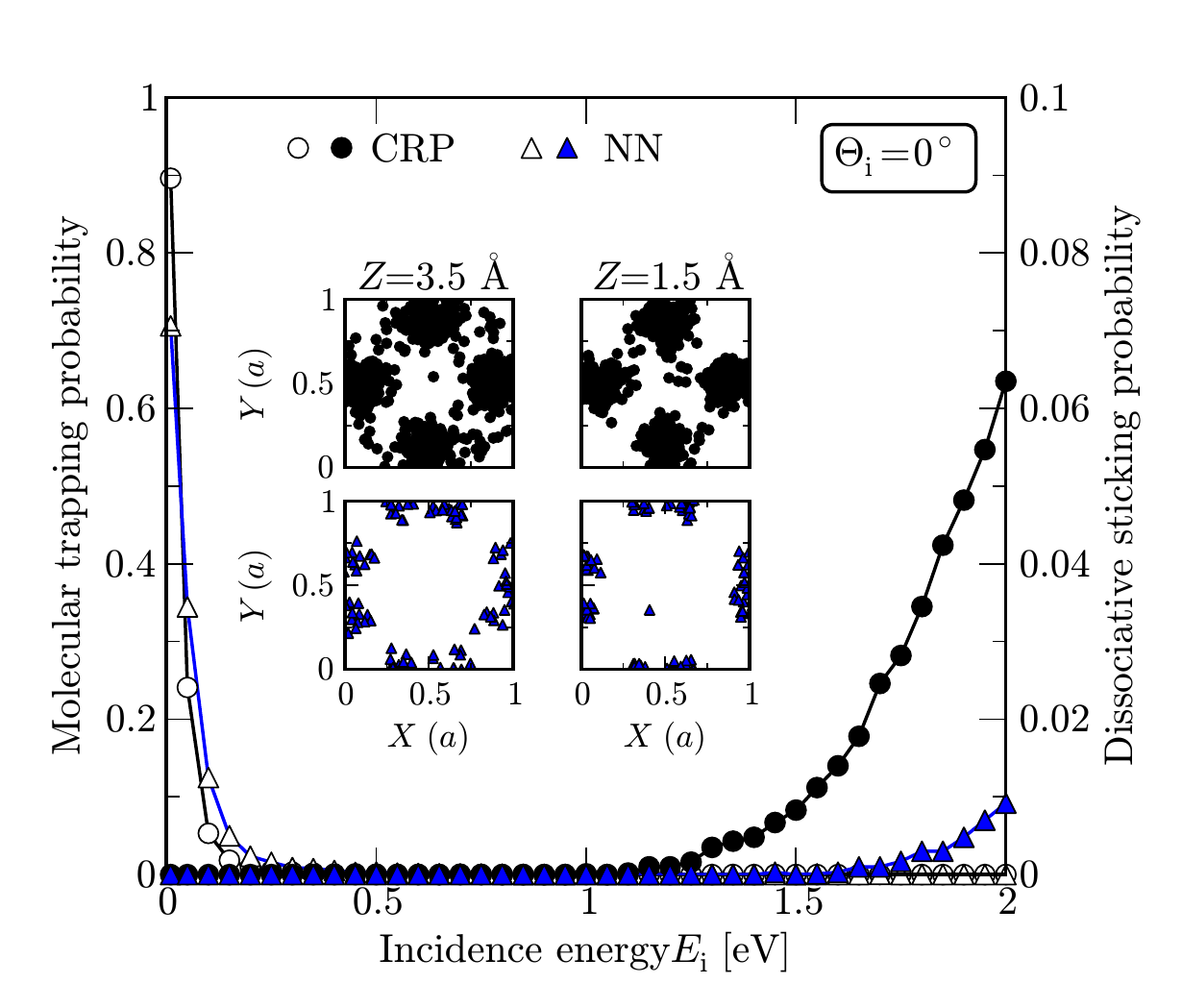}
\caption{Molecular trapping (open symbols) and dissociative sticking (filled symbols) probabilities of O$_2$ impinging at normal incidence $\Theta_i = 0^{\circ}$ on the Ag(100) surface as a function of the incidence kinetic energy $E_i$. The insets show the position of the molecular center over the unit cell, when the dissociating molecules are at $Z =3.5~ \mathrm{\AA}$ (left panels) and $Z =1.5\,\mathrm{\AA}$ (right panels) for an incidence energy of 2.0 eV. Note that the same coordinate system as in Fig.~\ref{fig:fig2} is used, i.e. surface atoms are located at the corners of the depicted unit cell. In all cases, circles and triangles correspond to data obtained for the CRP- and NN-PES representation, respectively.}
\label{fig:fig4} 
\end{figure}

The errors obtained with both approaches are {\em en par}, if not smaller than those reported in a number of preceding divide-and-conquer studies, and would generally be considered as reflecting faithful state-of-the-art PES representations. Still, such assessment (even though standard) is obviously only based on those points included in the {\em ab initio} data set, which in the present case does not include lower-symmetry points, see the discussion below. That this represents a major shortcoming is prominently illustrated by the different results obtained in the dynamical simulations on these two PESs summarized in Fig. \ref{fig:fig4}: These differences center on the dissociative adsorption dynamics of the system, whereas quite similar results are obtained for molecular trapping at low incidence energies. For the normal incidence case shown in Fig.~\ref{fig:fig4}, this specifically amounts to an onset of the dissociative sticking probability at about {1.05\,eV} in the CRP case, while the onset is delayed to {1.6\,eV} for the NN representation. With a monotonously rising dissociative sticking probability, this shift in the onset leads to much higher values in the CRP case at higher incidence energies, which e.g. with 0.06 at $E_i = 2.0\,\mathrm{eV}$ surpasses the equivalent NN probability by more than 500\%. The same trend is observed for various incidence angles tested, indicating a much higher reactivity of the CRP representation in general with the dissociative sticking probability already starting to rise at significantly lower incidence kinetic energies.

\subsection{Differences in reaction mechanism}\label{sec:results_dynamics}

In terms of absolute values for the sticking probability, the differences between the two approaches might not appear too worrisome at first sight. Due to the large number of trajectories employed in the averaging, they are, however, statistically significant. As such, the huge relative discrepancies question the divide-and-conquer approach in its very core regime inaccessible to the alternative direct AIMD ansatz, namely the quantitative determination of very low reaction probabilities. 

Since both PES representations are based on the exact same DFT data, the differences must derive from interpolation deficiencies that go unnoticed in the traditional quality indicators based on fitting errors. As a first step towards understanding the source of the problem and establishing protocols to overcome it, we focus on an analysis of the dissociation mechanism. Ideally, this will provide insight into which parts or topological features of the PES are central to the sticking coefficient. In this respect, the original work on the CRP PES by Alducin \textit{et al.} emphasized the role of high energy barriers of about 1.05\,eV between bridge and hollow sites at a distance of 1-2\,{\AA} above the surface \cite{Alducin2008}. They reflect molecules with low incidence energies, while molecules with a kinetic energy large enough to surmount them are subsequently capable of sufficiently approaching the surface to finally dissociate in the vicinity of the Ag(100) bridge sites. 

Meticulous searches for minimum energy paths performed via the Nudged Elastic Band (NEB) method within the present work, revealed such barriers to also exist along the most energetically favorable path traced towards dissociation within the NN representation. This suggests that the latter yields at least a qualitatively similar reaction mechanism as the CRP representation; a suspicion that we find fully confirmed by the trajectory data shown in the insets of Fig.~\ref{fig:fig4}. Analyzed is the O$_2$ center of mass position over the surface unit-cell of ultimately dissociating molecules as they first reach a surface distance of $Z = 3.5\,\mathrm{\AA}$ (left panels) and $Z=1.5\,\mathrm{\AA}$ (right panels) for an incidence energy of {2.0\,eV}. Indeed, in both PES representations the molecules predominantly accumulate around the bridge position, underscoring the relevance of this PES ``region'' for the dissociation mechanism in both representations. However, in the NN case (reflecting the lower dissociative sticking probability) this is a significantly smaller number of trajectories. Furthermore, by comparing the distributions of the trajectories in both insets in detail, one notices that in the CRP case dissociating trajectories are found to be homogeneously distributed within the vicinity of the bridge sites, whereas for the NN particularly in the direction towards the hollow sites no dissociating trajectories can be found. This indicates PES interpolation deficiencies in this area to be primarily responsible for the differing reaction probabilities obtained with the two representations.

\subsection{Picturing the PES ``landscape'': a visual comparison}\label{sec:results_2Dcuts}

\begin{figure}
\includegraphics[width=\colwidth]{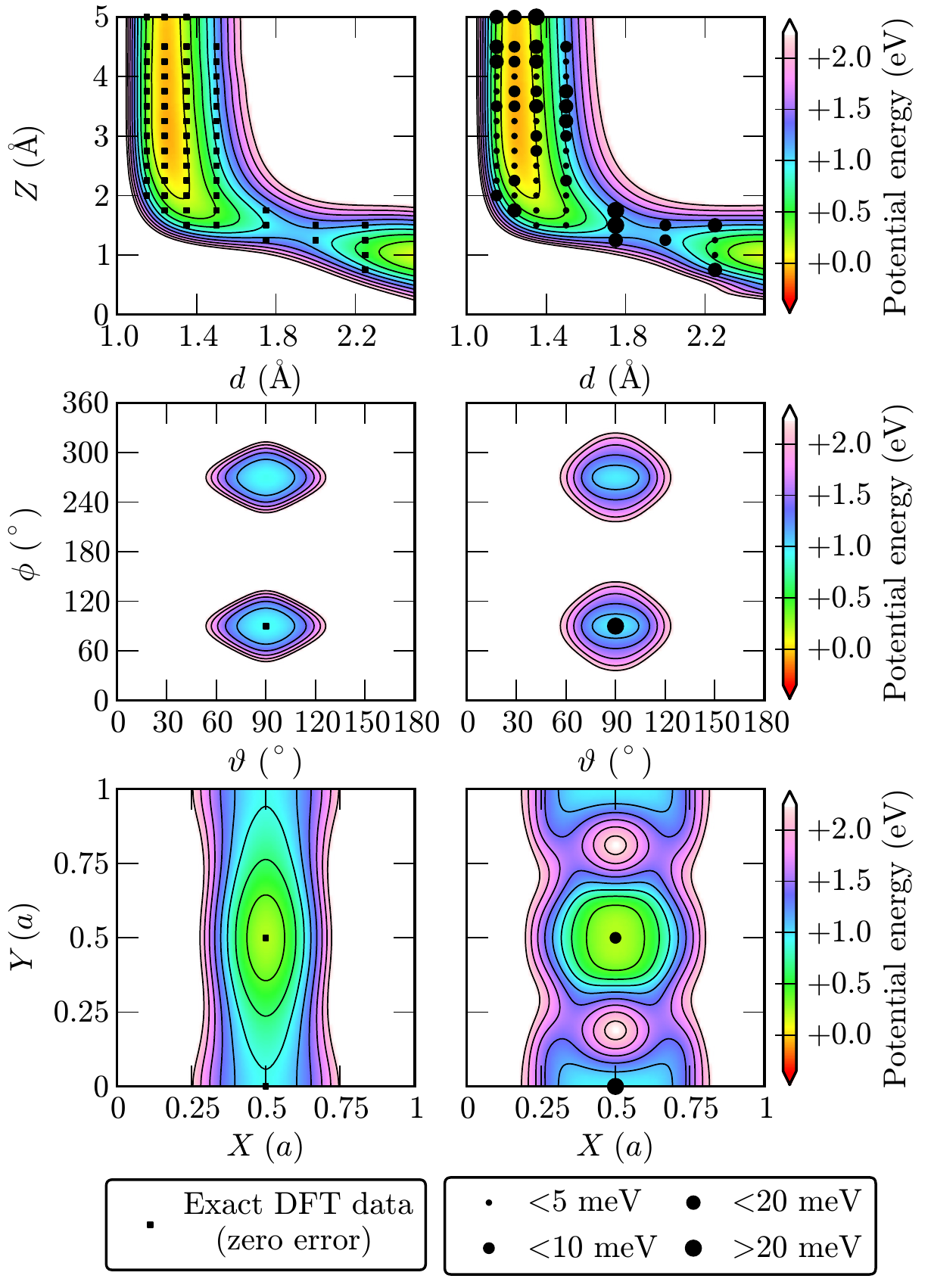}
\caption{Equivalent representations of the CRP (left) and NN (right) potential energy surfaces. All panels represent two-dimensional cuts through the 6D PESs: for $X = 0.5\,a$, $Y = 0$ and $\theta = \phi =90^\circ$ (top); for $X = 0.5\,a$, $Y = 0$, $Z = 1.5\,\mathrm{\AA}$ and $d =1.75\,\mathrm{\AA}$ (middle); for $Z = 1.5,\mathrm{\AA}$, $d =1.75~ \mathrm{\AA}$ and $\theta = \phi = 90^\circ$ (bottom). The black points outline the underlying DFT grid with the marker shape and size illustrating the corresponding errors of the interpolated representation as explained in the figure's legend.}
\label{fig:fig5}
\end{figure}

A most straightforward way to investigate if and how the two interpolation approaches give rise to different topologies in those PES regions that the preceding analysis identified as most influential for the dynamical behavior is to visually compare suitable two-dimensional (2D) cuts through the two 6D PES representations and include the actual {\em ab initio} data. Figure~\ref{fig:fig5} depicts three corresponding cuts: A ($d$-$Z$) elbow cut over the bridge site and for alignment of the molecular axis along the [001] direction ($\vartheta = \varphi = 90^{\circ}$), i.e. in the direction of the neighboring hollow sites; a ($\vartheta$-$\varphi$) angular plot over the bridge site and at a surface distance of 1.5\,{\AA}, i.e. somewhere at the $Z$ of the dissociation barriers; and a ($X$-$Y$) lateral corrugation plot at the same distance from the surface and for the most favorable angular orientation identified in the ($\vartheta$-$\varphi$) cut (that again points towards the hollow sites). Additionally shown are the coordinates of the actual DFT data grid in each 2D cut. By construction, the DFT values at these grid points are exactly reproduced by the interpolation technique employed in the CRP approach (cf. Sect.~\ref{sec:methodology_interpolation}). In contrast, within the NN representation they are only fitted and the marker size at the grid point reflects the corresponding error.

Consistent with the weight factors assigned upon training and with the initially discussed small global fitting errors, specifically the dynamically relevant low energy points are very well reproduced by the NN. At the high density of DFT data points in the ($d$-$Z$) elbow plot, this (or the exact reproduction of the DFT values in the CRP case) is enough to rather unambiguously determine the continuous representation within the corresponding plane, resulting in the remarkable resemblance of both plots shown in the upper panels of Fig.~\ref{fig:fig5}. This also holds for all other eleven elbows contained in the DFT data set (not shown). Despite the much sparser supporting grid (of in fact only five inequivalent DFT data points), the same still seems to be valid in the angular ($\vartheta$-$\varphi$) plane shown in the middle panels of Fig.~\ref{fig:fig5} (where only a single of the five DFT points falls within the depicted energy range). In the lateral ($X$-$Y$) plane, however, along which DFT sampling is noticeably ``poor'', both representations exhibit significant differences in the levels of corrugation and overall features of the PES landscape, with the NN PES featuring an overall more complex topology. Similar conclusions are drawn from a large number of 2D cuts analogous to those of Fig.~\ref{fig:fig5} for which (some) DFT data points are available. These cuts generally reflect a higher degree of ``creativity'' of the NN in describing the PES within areas that are sparsely supported by {\em ab initio} data. 

In principle, differences in the description of sparsely supported PES areas are not surprising, but in fact to be expected: If imagining the continuous PES to be like a carpet that has been nailed to the floor at specific points (by locally minimizing the RMSE with respect to the DFT data), ``bumps'' and ``folds'' can freely reveal themselves ``between'' the (more or less) fixed points within both approaches. That there are more such ``bumps'' and ``folds'' in the NN case is hereby not related to targeting the more corrugated $V_\mathid{6D}$ with the NN rather than targeting the smoother interpolation function $\mathcal{I}^\mathid{CRP}$ used in CRP. Similar to recent work by Ludwig and Vlachos \cite{Ludwig2007}, but including proper treatment of symmetry, we have also set an equivalent $\tilde{\mathcal{F}}^\mathid{NN}(\{\mathvec{Q}_i(\mathvec{R}_i^\mathid{cart},\mathvec{R}_i^\mathid{sph})\})$ function as the target for the global fitting in symmetry adapted coordinate space with the NN (cf. Sect.~\ref{sec:methodology}). Pronounced differences in the lateral dimensions, however, are still obtained for the different representations. Consequently, the disentanglement into independent interpolation steps employed in CRP (cf. Sect.\ref{sec:methodology_interpolation}) is instead identified as the cause for the altogether smoother CRP-PES. In particular, the Fourier interpolation in $(X,Y)$ does include physical assumptions about the PES behavior in these degrees of freedom which are absent in the NN case. The aforementioned mixed blessing of the latter is thus nicely exemplified: On the one hand, it can produce the observed ``creative'' PES representations. On the other hand, this can also allow to better capture more complex PES topologies that are correlationally induced by all six molecular coordinates and which are then beyond the flexibility of the functional form of the Fourier interpolation. The latter is e.g. reflected by the lower root mean square error of the NN test set described in Sect.~\ref{sec:results_rmse}.

Altogether, the 2D cuts presented in Fig. \ref{fig:fig5} nicely complement the understanding derived from the mechanistic analysis: The ($d$-$Z$) elbow cuts over the bridge site confirm that both PES representations yield similar barriers for this dissociation path that wants the molecular axis parallel to the surface and steered into a side-on orientation along the [001] direction. In contrast, the markedly different PES topologies encountered in the $XY$ plane suggest that the consequences on the sticking probabilities can be attributed to the influence of ``bumps'' and ``folds'' of both representations ``in between'' these elbows. We turn in the following therefore, to locating such topological PES features, and prominently the minima, which as we will show not only helps to further pinpoint and understand the source of the dynamical discrepancies, but also provides a general protocol to assess the PES interpolation beyond the level of fitting errors.

\subsection{PES topology: in search of local minima}\label{sec:results_minima}

\begin{figure}
\includegraphics[width=\colwidth]{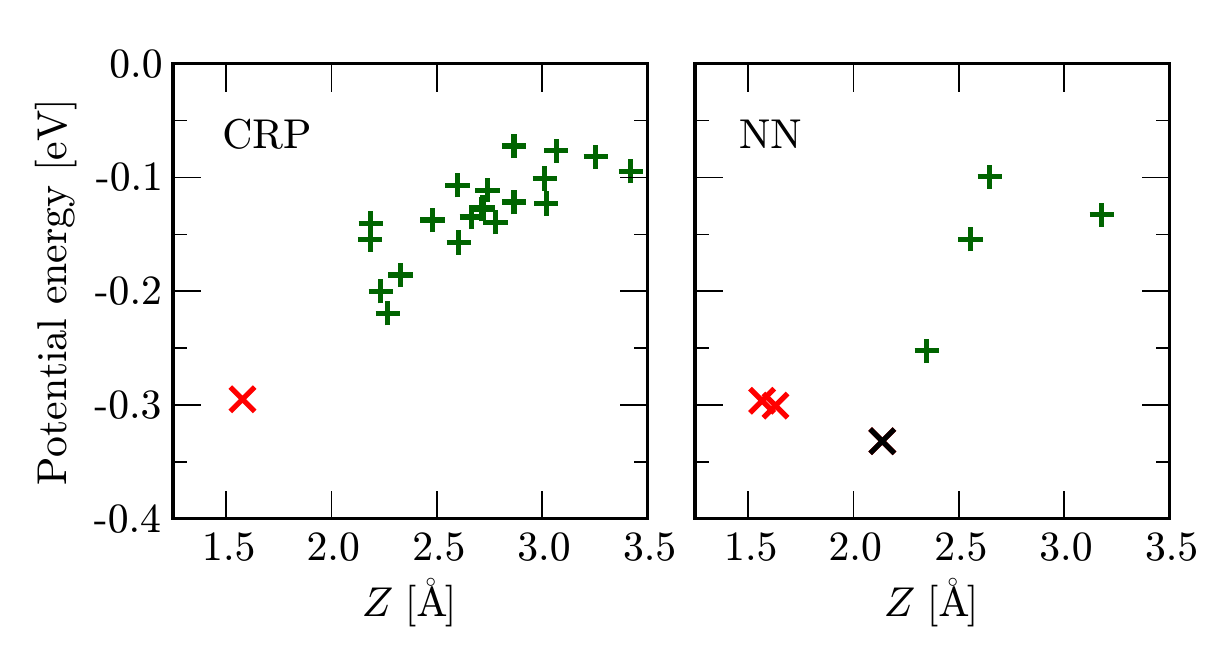}
\caption{Potential energy of the local minima configurations detected for the CRP (left) and NN (right) PES representations as a function of their corresponding distance from the surface, $Z$. In both cases, different symbols are used to conveniently set apart the low-energy structures (crosses) from the more `shallow' minima located further away from the surface (plusses). Note in particular, the NN global minimum (marked in black) that has no counterpart in the CRP representation.}
\label{fig:fig6}
\end{figure}

When applying the global search scheme described in Sect.~\ref{sec:methodology_minima} to both PES representations a significantly different and surprisingly high number of local minima is obtained. 21 and 7 bonded (with $V_{\mathid{6D}}<0$) molecular configurations are detected for the CRP and NN PESs, respectively. Some of these minima are rather shallow minima on the continuous representations, and thus not all of them are likely to correspond to physically meaningful adsorptive structures.

These minima are plotted in Fig. \ref{fig:fig6} as a function of their vertical distance $Z$ from the surface. A detailed analysis of trajectory ensembles (\emph{vide infra}) excludes the entrance channel as having a major influence on the dynamics by effectively steering the impinging molecules into orientations or impact sites of different reactivity within the two PES representations. We therefore suspect minima closest to the surface to have the biggest influence on the dissociation dynamics (if there is any), imagining them to be most effective for an activation of the molecular bond, i.e. transfer of energy into the $d$ coordinate. Clearly separated from the rest, only one such minimum is observed for the CRP representation at a surface distance of approximately $1.58\,\mathrm{\AA}$ and an adsorption energy of ca. -300 meV. This minimum is scrupulously reproduced within the NN representation, which, however, exhibits two additional minima (at $Z=1.63\,\mathrm{\AA}$ and $2.14\,\mathrm{\AA}$) in a similar energy range. Since one of these corresponds in fact to even the unique global minimum of the NN representation $M_\mathid{NN} = (X_\mathid{min} = 0.25~a, Y_\mathid{min} = 0.5~a, Z_\mathid{min} = 2.14\,\mathrm{\AA}, d_\mathrm{min} = 1.28\,\mathrm{\AA}, \vartheta_\mathid{min} = 90^\circ, \varphi_\mathid{min} = 90^\circ)$ in the irreducible wedge (cf. Fig.~\ref{fig:fig2}), we start by focusing our analysis on its influence.

\begin{figure}
\includegraphics[width=\colwidth]{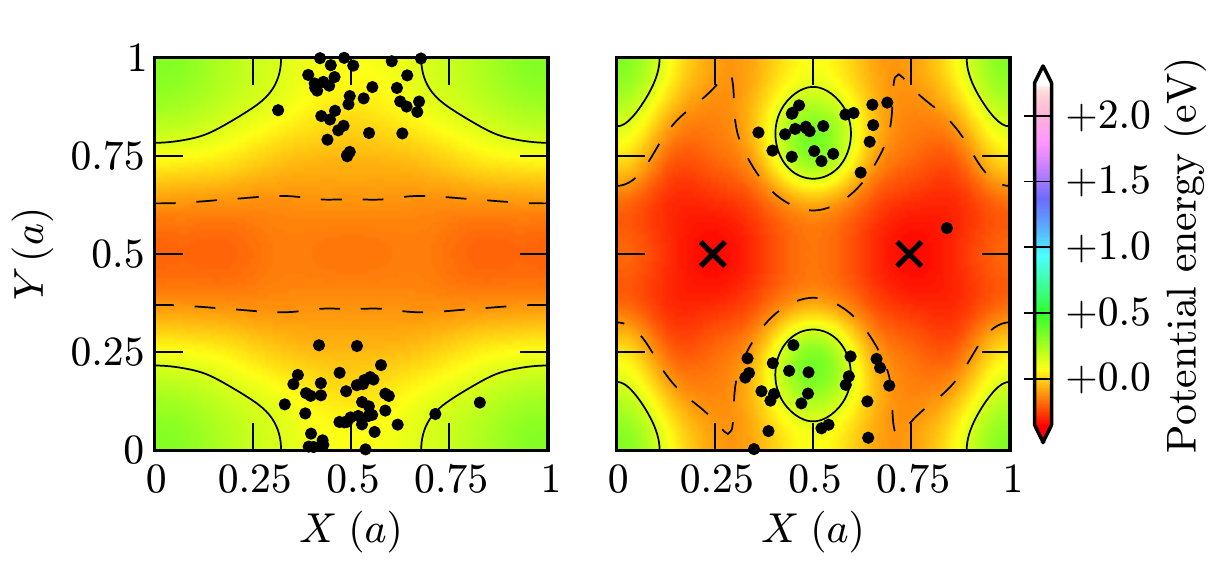}
\caption{Equivalent two-dimensional cuts of the CRP (left) and NN (right) PES representations similar to the lower panel of Fig.~\ref{fig:fig5}. The ($X$-$Y$) plots correspond to $Z_\mathid{min}$, $d_\mathrm{min}$, $\vartheta_\mathid{min}$ and $\varphi_\mathid{min}$, i.e. they are cuts along the lateral plane through the NN global minimum at ($X_\mathid{min} = 0.25~a$, $Y_\mathid{min} = 0.5~a$, $Z_\mathid{min} = 2.14\,\mathrm{\AA}$, $d_\mathrm{min} = 1.28\,\mathrm{\AA}$, $\vartheta_\mathid{min} = 90^\circ$, $\varphi_\mathid{min} = 90^\circ$) in the irreducible wedge (cf. Fig.~\ref{fig:fig2}).
Note that due to the symmetry of the angular coordinates $(\vartheta_\mathid{min}, \varphi_\mathid{min})$, a horizontal and vertical mirror plane perpendicular to the surface through the hollow site (i.e. the center of the panels) is preserved from the clean Ag(100) surface symmetry within these cuts. Consequently, there is also a symmetry equivalent counterpart of the NN global minimum at $(\tilde{X}_\mathid{min} = a - X_\mathid{min},Y_\mathid{min},Z_\mathid{min}, d_\mathrm{min} \vartheta_\mathid{min},\varphi_\mathid{min})$. Both minima are marked by black crosses. The black circles correspond to trajectory data and show the respective positions of the molecular center in this cut plane (folded back into the unit cell) for molecules that pass closely through this cut plane, i.e. $(Z,d,\vartheta,\varphi) \in [Z_\mathid{min},d_\mathid{min} \pm 0.05\,\mathrm{\AA}, \vartheta_\mathid{min} \pm 30^{\circ},\varphi_\mathid{min} \pm 20^{\circ}]$. Other than in the inset of Fig.~\ref{fig:fig4}), only equivalent trajectories starting from the same initial conditions with an incidence energy of 2.0\,eV are included that dissociate on the CRP (left) but are reflected on the NN PESs (right).}
\label{fig:fig7}
\end{figure}

\begin{figure}
\includegraphics[width=\colwidth]{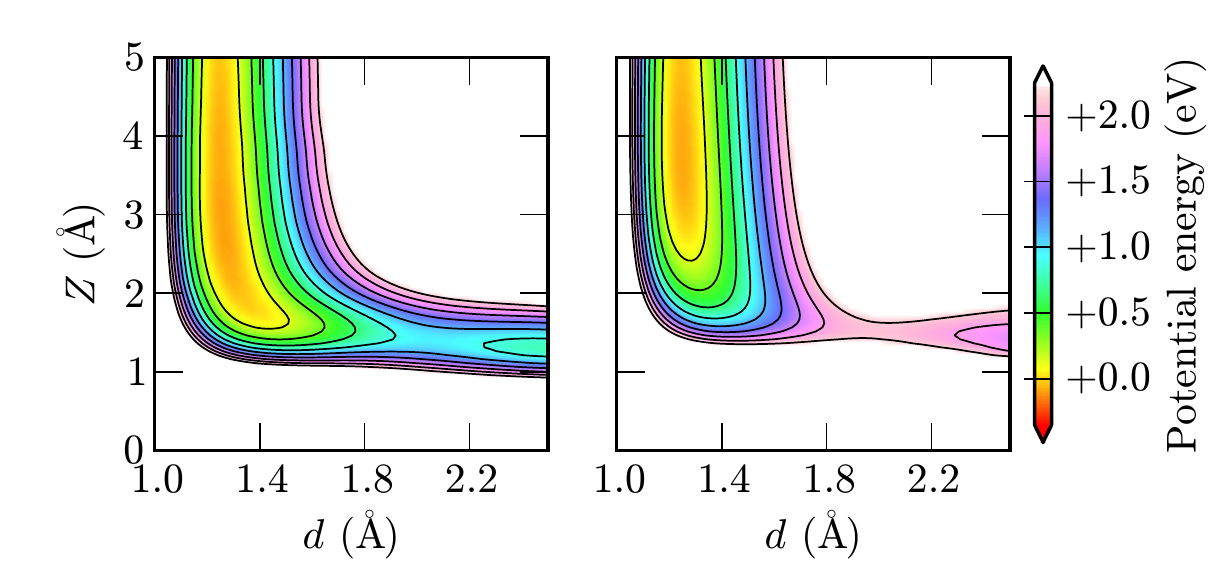}
\caption{Equivalent two-dimensional cuts of the CRP (left) and NN (right) PES representations similar to the upper panel of Fig.~\ref{fig:fig5}. The ($d$-$Z$) elbow plots are shown for $X_\mathid{min}$, $Y_\mathid{min}$, $\vartheta_\mathid{min}$ and $\varphi_\mathid{min} - 90^\circ$, i.e. the coordinates of the ``bumps'' identified in Fig.~\ref{fig:fig7} backfolded in the irreducible wedge (see text).}
\label{fig:fig8}
\end{figure}

Returning to the examination of 2D cuts through the PES representations, the upper part of Fig.~\ref{fig:fig7} visualizes similarly large differences in the lateral degrees of freedom in the vicinity of this NN global minimum as discussed already in Sect.~\ref{sec:results_2Dcuts}. In order to analyze the influence on the dynamics, we have also included $(X,Y)$ coordinates of molecules with $E_i = 2\,\mathrm{eV}$ that pass through the close vicinity of this plane as detailed in the figure's caption. Other than in the inset of Fig.~\ref{fig:fig4}, we now concentrate on all those trajectories that dissociate on the CRP PES and whose equivalent counterparts (i.e. starting from the exact same initial conditions) are reflected in the case of the NN. This particular choice of depicted trajectories (CRP dissociating and NN reflected) indicates that the dynamical discrepancies are not to be traced back to effects of molecular steering, but are rather due to the different topology of the PES representations in regions close to the surface. Still, this ensemble does not pass through $M_\mathid{NN}$ itself, such that the minimum itself can not directly be held responsible for their radically different outcome. Instead, the trajectory distribution is more or less centered around repulsive ``bumps'' only existing in the NN representation shown in Fig.~\ref{fig:fig7} at $(0.5~a, 0.5~a \pm 0.25~a, Z_\mathid{min}, d_\mathrm{min}, \vartheta_\mathid{min}, \varphi_\mathid{min})$. Both ``bumps'' can be identified with a symmetry equivalent representative $B_\mathid{NN} = (X_\mathid{min}, Y_\mathid{min}, Z_\mathid{min}, d_\mathrm{min}, \vartheta_\mathid{min}, \varphi_\mathid{min}-90^\circ)$ in the irreducible wedge (cf. Fig.~\ref{fig:fig2}) with $V_\mathid{6D}^\mathid{NN}(B_\mathid{NN}) \approx +0.3~\mathrm{eV}$. This motivates picturing the NN PES representation like in Sect.~\ref{sec:results_2Dcuts} at this ``suspicious location'' within the close vicinity of $M_\mathid{NN}$ also in the other degrees of freedom. Fig.~\ref{fig:fig8} compares CRP and NN elbow plots for a corresponding molecular configuration. While the CRP PES still shows a dissociation path with a barrier of about $1\,\mathrm{eV}$ equivalent to the one directly encountered at the bridge site (cf. upper part of Fig.~\ref{fig:fig5}), its NN counterpart is by more than $1\,\mathrm{eV}$ steeper. Very similar trends are also observed for other elbows corresponding to different angular orientations of the molecule. For molecules therefore with a kinetic energy just sufficient to traverse the minimum energy path all these configurations would already lead to dissociation within the CRP PES yet be repelled by the much higher barriers on the NN PES.

This fully rationalizes the observed differences in reaction probabilities shown in Fig.~\ref{fig:fig4} in terms of the differing degree of creativity for the CRP and NN pictures discussed previously: While the DFT data grid is dense enough to yield the same qualitative dissociation mechanism in both representations, additional ``bumps'' in the NN representation yield a much smaller lateral extent of this dominant dissociation pathway and thereby lead to a significantly reduced dissociation probability. Additional single-point DFT calculations performed in these regions indeed reveal large representation errors of $\approx 240~\mathrm{meV}$ and confirm the interpretation of the ``bumps'' as unphysical products of the NN fit. Correspondingly large errors associated with the NN local minima (of which a detailed account may be found in supplementary materials) further suggest that the additional ``bumps'' and ``folds'' tend to go hand in hand, thus also resulting in a differing amount of extremal points on the PES, as indicated in the present case by the differing number of low-energy minima identified on the NN and CRP PESs.

In a more general context however, whether such topological features are real or spurious -- and correspondingly which PES representation is to be trusted more -- cannot be answered {\em a priori}, but only by iteratively refining the DFT data grid and thus successively reducing the interpolation freedom. Preceding any detailed dynamical simulations, a static global minimum search as performed here appears therefore as a suitable and computationally undemanding general protocol to assess the interpolation quality beyond the level of an arbitrarily chosen test set of \textit{ab initio} data: Particularly PES regions around the identified minima should be supported by a sufficiently dense DFT data grid - if not in the original data set, then at least through iterative refinement. While not at all being a conceptual limitation of the central idea behind CRP, this is more conveniently carried out within the present NN implementation at the moment and also more important for the latter at least in the present case.

\section{Conclusions \& Outlook}\label{sec:conclusions}

In summary, we have presented a systematic comparison of continuous potential energy surface (PES) representations commonly used in gas-surface dynamics within the divide-and-conquer approach. Employing the exact same set of \textit{ab initio} data \cite{Alducin2008} for the showcase system O$_2$ on Ag(100) such six-dimensional PES representations according to two prevalent schemes, namely the Corrugation-Reducing Procedure (CRP) and Neural Networks (NNs), were found to yield largely differing dissociative sticking probabilities, despite flawlessly passing commonly employed representation quality checks. In particular for statistical quantities with low probabilities (such as the dissociative sticking probability focused on here) direct {\em ab initio} molecular dynamics is unlikely to replace the divide-and-conquer approach in the near future. This being the case, the reported results are particularly irritating and dictate the development of protocols to more reliably assess and refine the representation quality.

To this end and complementing an evaluation through explicit dynamical trajectory simulations we emphasize the value of a static evaluation of the obtained PES. In the spirit of Eyring and Polanyi, this concerns foremost the imperative visualization along suitable 2D cuts. In the present case this allowed to trace the differences in reaction probabilities back to systematic differences of the two approaches in areas that are only sparsely sampled by DFT input data. With the latter conventionally concentrated in $(Z,d)$ elbow cuts, this applies especially to lateral degrees of freedom, where we found the NN PES representation to exhibit a much more complex topology with additional ``bumps'' and ``folds''. In contrast, the physical assumptions embodied in the Fourier interpolation that is used for these lateral degrees of freedom within the decoupled, at most two-dimensional interpolation steps employed in prevalent CRP versions necessarily yield a much smoother description.

For the showcase system, the additional ``bumps'' were shown to lead to a significantly reduced lateral extent of the dominant dissociation pathway and thereby to the much reduced dissociation probability obtained with the NN PES representation. Both here and in general it is {\em a priori} impossible to judge which of the two strategies yields the more reliable global representation: The flexibility of the NN approach, directly targeting the six-dimensional PES and its full topology, can allow to foresee/reproduce PES features which the CRP approach with its decoupling strategy systematically excludes. This can extend to (technical) inconsistencies of the underlying \textit{ab initio} data, which spin-polarized density-functional calculations for systems -- like the present one -- that exhibit a spin transition are particularly prone to. On the other hand, this high flexibility brings along the risk of a high ``creativity'' and spurious artifacts in sparsely sampled PES regions.

Ultimately, the only reliable answer as to the more faithful representation can be given by explicit {\em ab initio} molecular dynamics simulations. Yet, this is precisely what the divide-and-conquer approach tries to avoid. Alternatively, one could always argue to construct ''better'' {\em ab initio} training sets, e.g. including more data at low-symmetry sites. In an increasing dimensionality of the problem this becomes a more and more delicate task with unclear outcome though. In our view, a more controlled approach centers on an iterative refinement of the input data density in sparsely sampled PES areas and in particular in those that are relevant for the dynamics. In understanding the role of additional ``bumps'' and ``folds'' created by the NN approach, we propose that a global minima search provides a most helpful (and practically manageable) tool to guiding such a refinement endeavor. In a first place one would simply want a continuous PES representation to yield a qualitatively correct topology and therefore knowledge of the representation's low-energy minima always comprises a good starting point. Much more, however, the added value of locating low-energy regions of the PES lies in their general likelihood of influencing the system's dynamics. Even if this is not the case -- as for the dissociative sticking studied here -- the aforementioned ``bumps'' and ``folds'' (aka topological features) necessarily go hand in hand. A global search for the latter is in this respect a useful (and numerically undemanding) complement to a dynamical trajectory analysis on the PES representation, which would tediously have to be repeated for different incidence energies or angles. The knowledge of the minima can then also directly be employed to steer a visual PES exploration by focusing on 2D cuts that go through their positions. As we have demonstrated here, such a look in the vicinity of these minima can reveal ``bumps'', which in case of sparse DFT data sampling around the latter quickly raises suspicions about them being artifacts of the continuous representation. In this manner, regions of the PES are identified for which adding further data points and pinpointing the PES ``carpet'' will most effectively increase the reliability of the continuous representation. This holds in particular when not having a second such representation at hands for comparison, which after all corresponds to the usual working situation.

% include supplementary material for publication on arXiv


\section*{Supplementary material}

# Supplementary materials

The present section provides a detailed presentation of the optimized configurations as generated by the stochastic searches for local minima on the six-dimensional $O_2/Ag(100)$ PES, which was continuously represented via both the Corrugation-Reducing Procedure (CRP) and a Neural Network (NN) fit (Tables 1 and 2, respectively). Listed are the optimized molecular geometries and the corresponding adsorption energies as extracted by the interpolated resp. fitted PES representations and as re-calculated within DFT.

## 1. CRP local minima

Table 1: List of optimized configurations as generated by the stochastic search for local minima of the $O_2/Ag(100)$ CRP PES representation. Entries are listed and numbered in order of increasing optimized energy. $V_{6D}^{CRP}$ and $V_{6D}$ are the adsorption energies of the corresponding molecular configuration as obtained within the six-dimensional CRP-interpolated PES and as re-calculated within the same VASP set-up employed by Alducin *et al.* in earlier work [J. Chem. Phys. **129** (2008) 224702], respectively. The differences between the two aforementioned quanitites are given in the subsequent row and denote the (single-point) errors of the interpolation. All molecular geometries are expressed in spherical coordinates where the lateral coordinates $(X, Y)$ are given in units of the theoretical surface lattice constant $a$=2.95 Å.

| Minimum no.        | 1      | 2      | 3      | 4      | 5      | 6      |
|--------------------|--------|--------|--------|--------|--------|--------|
| $V_{6D}^{CRP}$ (eV) | -0.295 | -0.219 | -0.200 | -0.186 | -0.157 | -0.155 |
| $V_{6D}$ (eV)      | -0.345 | -0.266 | -0.271 | -0.250 | -0.151 | -0.222 |
| Error (eV)         | +0.050 | +0.047 | +0.071 | +0.064 | -0.006 | +0.067 |
| $X$ ($a$)          | 0.50   | 0.50   | 0.17   | 0.50   | 0.50   | 0.51   |
| $Y$ ($a$)          | 0.50   | 0.00   | 0.50   | 0.25   | 0.13   | 0.33   |
| $Z$ (Å)            | 1.58   | 2.27   | 2.23   | 2.33   | 2.60   | 2.18   |
| $d$ (Å)            | 1.42   | 1.29   | 1.29   | 1.28   | 1.27   | 1.29   |

Table 1 – continued from previous page

| | | | | | | |
|---|---|---|---|---|---|---|
| $\theta$ (degrees) | 90.0 | 90.0 | 90.0 | 90.2 | 137.3 | 90.9 |
| $\phi$ (degrees) | 180.0 | 180.0 | 90.0 | 357.5 | 270.0 | 160.0 |
| | | | | | | |
| Minimum no. | 7 | 8 | 9 | 10 | 11 | 12 |
| | | | | | | |
| $V_{6D}^{CRP}$ (eV) | -0.140 | -0.139 | -0.137 | -0.135 | -0.128 | -0.126 |
| $V_{6D}$ (eV) | -0.193 | -0.164 | -0.185 | -0.177 | -0.189 | -0.168 |
| Error (eV) | +0.053 | +0.025 | +0.048 | +0.042 | +0.061 | +0.042 |
| | | | | | | |
| $X$ ($a$) | 0.51 | 0.17 | 0.41 | 0.20 | 0.32 | 0.44 |
| $Y$ ($a$) | 0.51 | 0.35 | 0.50 | 0.34 | 0.30 | 0.38 |
| $Z$ (Å) | 2.19 | 2.78 | 2.48 | 2.67 | 2.71 | 2.72 |
| $d$ (Å) | 1.29 | 1.26 | 1.27 | 1.27 | 1.26 | 1.26 |
| $\theta$ (degrees) | 90.2 | 52.7 | 87.9 | 113.8 | 73.9 | 78.5 |
| $\phi$ (degrees) | 225.3 | 24.3 | 360.0 | 218.8 | 72.5 | 30.4 |
| | | | | | | |
| Minimum no. | 13 | 14 | 15 | 16 | 17 | 18 |
| | | | | | | |
| $V_{6D}^{CRP}$ (eV) | -0.123 | -0.122 | -0.111 | -0.107 | -0.101 | -0.094 |
| $V_{6D}$ (eV) | -0.100 | -0.120 | -0.128 | -0.162 | -0.187 | -0.152 |
| Error (eV) | -0.023 | -0.002 | +0.017 | +0.055 | +0.086 | +0.058 |
| | | | | | | |
| $X$ ($a$) | 0.35 | 0.31 | 0.50 | 0.50 | 0.12 | -0.21 |
| $Y$ ($a$) | 0.35 | 0.31 | 0.36 | 0.50 | -0.12 | -0.11 |
| $Z$ (Å) | 3.02 | 2.87 | 2.74 | 2.60 | 3.01 | 3.42 |
| $d$ (Å) | 1.26 | 1.26 | 1.27 | 1.28 | 1.26 | 1.25 |
| $\theta$ (degrees) | 149.1 | 46.3 | 167.1 | 0.0 | 115.2 | 119.4 |
| $\phi$ (degrees) | 225.1 | 45.0 | 270.0 | 28.8 | 135.7 | 43.8 |
| | | | | | | |
| Minimum no. | 19 | 20 | 21 | | | |
| | | | | | | |
| $V_{6D}^{CRP}$ (eV) | -0.082 | -0.077 | -0.072 | | | |
| $V_{6D}$ (eV) | -0.127 | -0.120 | -0.070 | | | |
| Error (eV) | +0.045 | +0.043 | -0.002 | | | |

Table 1 – continued from previous page

| | | | |
|---|---|---|---|
| $X$ ($a$) | 0.22 | 0.44 | 0.00 |
| $Y$ ($a$) | -0.19 | 0.15 | -0.00 |
| $Z$ (Å) | 3.25 | 3.01 | 2.87 |
| $d$ (Å) | 1.25 | 1.25 | 1.26 |
| $\theta$ (degrees) | 90.2 | 103.7 | 90.1 |
| $\phi$ (degrees) | 225.8 | 658.7 | 135.9 |

## 2. NN local minima

Table 2: List of optimized configurations as generated by the stochastic search for local minima of the $O_2$/Ag(100) NN PES representation. Entries are listed and numbered in order of increasing optimized energy. $V_{6D}^{NN}$ and $V_{6D}$ are the adsorption energies of the corresponding molecular configuration as obtained within the six-dimensional NN-fitted PES and as re-calculated within the same VASP set-up employed by Alducin *et al.* in earlier work [J. Chem. Phys. **129** (2008) 224702], respectively. The differences between the two aforementioned quanitites are given in the subsequent row and denote the (single-point) errors of the fit. All molecular geometries are expressed in spherical coordinates where the lateral coordinates ($X$, $Y$) are given in units of the theoretical surface lattice constant $a$=2.95 Å.

| Minimum no. | 1 | 2 | 3 | 4 | 5 | 6 | 7 |
|---|---|---|---|---|---|---|---|
| $V_{6D}^{NN}$ (eV) | -0.332 | -0.301 | -0.296 | -0.252 | -0.154 | -0.133 | -0.099 |
| $V_{6D}$ (eV) | -0.243 | +0.008 | -0.346 | -0.232 | -0.089 | -0.146 | -0.160 |
| Error (eV) | -0.089 | -0.309 | +0.050 | -0.020 | -0.065 | +0.013 | +0.061 |
| $X$ ($a$) | 0.50 | 0.59 | 0.50 | 0.00 | 0.26 | 0.62 | 0.50 |
| $Y$ ($a$) | 0.25 | 0.42 | 0.50 | 0.39 | 0.06 | 0.21 | 0.50 |
| $Z$ (Å) | 2.14 | 1.63 | 1.57 | 2.35 | 2.56 | 3.18 | 2.65 |

Table 2 – continued from previous page

| | | | | | | | |
|---|---|---|---|---|---|---|---|
| $d$ (Å) | 1.28 | 1.37 | 1.43 | 1.28 | 1.27 | 1.25 | 1.27 |
| $\theta$ (degrees) | 90.0 | 90.1 | 90.0 | 91.6 | 83.8 | 165.7 | 0.1 |
| $\phi$ (degrees) | 180.1 | 58.0 | 180.0 | 90.0 | 258.3 | 134.7 | 6.3 |



\end{document}